*Universal emission intermittency in quantum dots, nanorods, and nanowires*

*Pavel Frantsuzov[1], Masaru Kuno[2], Boldizsár Jánko[1,4] and Rudolph A. Marcus[3]*

University of Notre Dame, Notre Dame, Indiana 46556, [1]Department of Physics, [2]Department of Chemistry and Biochemistry; [3]Noyes Laboratory of Chemical Physics, California Institute of Technology, Pasadena, California 91125, [4]Materials Science Division, Argonne National Laboratory, Argonne, Illinois 60439.

Virtually all known fluorophores exhibit mysterious episodes of emission intermittency. A remarkable feature of the phenomenon is a power law distribution of on- and off-times observed in colloidal semiconductor quantum dots (QDs), nanorods, nanowires and some organic dyes. For nanoparticles the resulting power law extends over an extraordinarily wide dynamic range: nine orders of magnitude in probability density and five to six orders of magnitude in time. Exponents hover about the ubiquitous value of -3/2. Dark states routinely last for tens of seconds, which are practically forever on quantum mechanical time scales. Despite such infinite states of darkness, the dots miraculously recover and start emitting again. Although the underlying mechanism responsible for this phenomenon remains a mystery and many questions persist, we argue that substantial theoretical progress has been made.

### Introduction

Few problems of early quantum mechanics remain unsolved today. Fluorescence intermittency is one exception. At the dawn of modern quantum mechanics, Niels Bohr predicted "quantum jumps" of electrons between discrete energy levels of atoms and molecules. Such jumps were observed directly during the 1980s with the advent of single ion traps [1]. An ongoing series of experiments brought spectacular progress to single molecule imaging [2], raising more questions than answers about these emission "jumps".

Basically all known types of fluorophores studied to date exhibit fluorescence intermittency. They include single molecules[3-8], fluorescent proteins[9], polymer segments[10], semiconductor nanoparticles[11-28], nanorods[29] and even nanowires (NWs).[30-32] Such jumps, where the fluorophore literally stops emitting light under continuous excitation, are very different from those predicted by Bohr. Rather than occurring on the micro- to millisecond timescales, they last seconds and even minutes. These experimentally observed off-times are vastly longer than any timescale one would retrieve from a standard quantum mechanical model: The "dark" state lives practically forever. After these long periods of darkness, the fluorophore eventually emerges into an emitting state. This would be impossible if off periods were simply permanent photobleaching events.

Even more intriguing is the statistics of the intermittency. Whereas Bohr would have postulated exponential distributions of "on"-times and "off"-times, universal power law probability densities are actually observed [12-14]. For colloidal QDs, this power law extends over an extraordinarily wide range that spans nine orders of magnitude in probability density and five to six orders of magnitude in time. This phenomenon is remarkable for many reasons. First, the experimentally observed distribution refuses to yield a time scale. Second, universality on a lesser scale has revolutionized our understanding about phase transitions [33]. Finally, such statistics arise from the behavior of *single* fluorophores, not an ensemble. Given that single organic molecules also blink and exhibit near-identical power law blinking kinetics [3-8], it is tempting to speculate that such striking similarity is not an accident. We therefore ask whether there exist unrecognized points of commonality between seemingly disparate entities such as molecules and semiconductor nanostructures. While the underlying mechanism for answering such questions remains a mystery, we argue in this perspective that many, but not all, key experiments have already been conducted and that substantial theoretical progress has been made.

### Key features of blinking

Figure 1 shows representative examples of blinking in QDs and NWs. An emission "trajectory" from both systems is provided, plotting emission intensity as a function of time. In both cases the intensity fluctuates. When a threshold is used to distinguish on- from off-states, second-to-minute off-times are apparent.

In principle, plotting both on-time and off-time probability densities on a semi-logarithmic plot enables one to extract characteristic rates for turning the fluorophore on or off. However, QDs, molecules and even NWs exhibit power law (i.e. scale-free) kinetics, indicating that the rate must be distributed over many timescales. A log-log plot demonstrates this, showing linear behavior for both on- and off-time distributions over many decades in probability density and time (QD: Figure 1 c,d; NW: Figure 1 e,f). Even more impressive is that the same power law appears in molecular systems[3-8] with linearity spanning up to 5 (7) decades in time (probability density).[5]

Let us now summarize several key facts learned from the first decade of observing QD blinking. Any self-consistent and comprehensive theory of fluorescence intermittency must account for the following:

**(a) The existence of power laws from a threshold analysis:** Blinking occurs over a wide range of timescales (μs to minutes). The shortest and longest times are currently set by experimental limitations. Distributions $P(t_{on})$, and $P(t_{off})$ can be fit to power laws of the form $t^{-m_{on/off}}$ where $m_{on/off}$ ranges from 1.2 to 2.0 [34,35]. Truncation times ("cutoffs") were discovered in on-time distributions [14]. Such cutoffs occur on the second timescale and may represent a competing physical process which interrupts power law blinking. A corresponding off-time cutoff in QDs has not been reported, although it is speculated to occur on an hour timescale [20]. Figures 1 e,f show examples of such cutoffs in recent NW data.

**(b) The existence of power law power spectral densities:** Pelton and Guyot-Sionnest[15] have demonstrated that the power spectrum of the ensemble QD emission exhibits a power law of the form $p(f) \sim f^{-\alpha} (\alpha \sim 1)$. This was also confirmed in the case of single

QDs. Furthermore, a kink in the power spectral density was recently observed with slopes reverting from $\approx 1$ at low frequencies to $\approx 2$ above 100 Hz [16].

*(c) The light-driven nature of the blinking process***:** Intermittency is light-induced, as indicated by experiments revealing statistical "aging" of emission trajectories [36]. The ensemble emission intensity decays under continuous excitation and recovers in the dark [20]. In existing experiments, the on-time distribution cut-off is inversely proportional to the excitation intensity.[24,37]

*(d) A general lack of temperature dependence:* On/off power law slopes are generally temperature independent between 10 K and 400 K [12-14]. This has led some to speculate tunneling or another temperature independent physical processes at play. However, a weak temperature sensitivity of on-time cutoffs has been observed.[14]

*(e) A connection to spectral diffusion:* Neuhauser and Bawendi have suggested that blinking is connected to another ubiquitous single molecule phenomenon: spectral diffusion[21]. Large shifts in the spectrum coincide with equally rare jumps of the intensity. This suggests a direct correlation between spectral diffusion and emission intermittency through the redistribution of charges on or nearby the QD surface.

*(f) A continuous distribution of emission intensities and excitation lifetimes:* Typical fluorescence intensity trajectories from single QDs do not mimic random telegraph noise[1]. Unique "on" and "off" levels are not seen. Instead, many intensity levels exist. Schlegel and Mews [18] have since found that such intensity fluctuations are correlated with changes in the QD emission lifetime. Additional studies [17,19] have confirmed this and have, in turn, shown through simultaneous emission quantum yield measurements that non-radiative recovery rate fluctuations are connected to intensity variations. Furthermore, Sher et al. [38] have observed lifetimes much longer than the usual radiative lifetime of ~20 ns with a power-law distribution towards longer times.

*(g) A sensitivity to electric fields:* Cichos and co-workers have found a sensitivity of on-time/off-time power law slopes to the dielectric permittivity of the surrounding [34]. Recently, Barbara and co-workers have also demonstrated that the emission from single QDs is modulated by externally applied electric fields [28]. The sign of the modulation varies slowly in time suggesting that blinking is connected to changes in local electric fields stemming from charges on the QD surface.

### *Theoretical models*

From the above list of key experimental findings, constraints on any comprehensive theoretical model, even if phenomenological, appear formidable. Most current blinking models can broadly be categorized into one of two groups: a) those that focus solely on analyzing power law statistics [39,40] and b) those that postulate a physical mechanism for intermittency [13,14,36,41-44]. We focus here on the latter models and provide representative cartoons in Figure 2.

The first QD blinking model was developed by Efros and Rosen [41]. This is an extremely successful model, and despite a few shortcomings, still lies at the heart of conventional wisdom behind explaining fluorescence intermittency. Within this model, QDs are thought to undergo Auger ionization events under photoexcitation. Electrons are ejected from the dot to surrounding acceptor-like states. This leaves behind a positively charged QD. Subsequent electron-hole pairs experience rapid Auger-like nonradiative relaxation to the ground state, quenching any emission and thereby rendering the particle dark. This process continues until the QD is neutralized [45,46].

The Efros/Rosen model provided the first intuitive picture for blinking, and became a major step towards understanding the phenomenon. However, a key problem remains: In sharp contrast with experimental findings, it predicts characteristic on/off rates and corresponding exponential on-time/off-time distributions. To circumvent these limitations, a series of modifications have been proposed. They include:

1. **Multiple trap model:** (**Figure 2b**) Verberk and Orrit [42] assume the existence of multiple electron traps near the QD. Due to a static distribution of trapping and de-trapping rates, varying with distance and/or trap depth, power law off-time distributions **(a)** are naturally obtained. Furthermore, this model readily explains the dependence of off-time power law slopes with the dielectric properties of the environment [34]. Finally, the lack of temperature dependence **(d)** can be explained through a tunneling process.

2. **Spectral diffusion model**: (**Figure 2c**) Shimizu and Bawendi [14] hypothesized a resonant tunneling mechanism where the diffusion of the environment's acceptor energy level causes power law distributed on-to-off and off-to-on kinetics. Tang and Marcus [43] later developed this by assuming spectral diffusion of both QD and acceptor state energies. A key prediction is a change in the slopes of both on-time and off-time power-laws from 3/2 at long times to 1/2 at short times. Interestingly, this has recently been corroborated by power spectral density experiments conducted by Pelton [16].

3. **Spatial diffusion model:** (**Figure 2d**) Margolin and Barkai [36] suggested that any ejected electron performs a 3D diffusion in space about the QD prior to its return. While the model naturally predicts a $t^{-3/2}$ distribution of off-times, deviations (cf. **(a)**) from the -1.5 exponent require the introduction of anomalous diffusion processes. Furthermore, a finite probability exists for the carrier to never return to the QD, leading to permanently dark dots in the limit of long observation times.

4. **Fluctuating barrier model**: (**Figure 2e**) Kuno and Nesbitt [13] have alternatively suggested a model where emission intermittency involves fluctuations in the height or width of a tunneling barrier between an electron within the QD and an external trap state. Furthermore, during the tunneling process, the local environment was postulated to change such that between each off-to-on or on-to-off transition the tunneling barrier would differ.

These proposals were successful in bringing the Efros-Rosen theory in line with many, but not all experimental constraints. In particular, a key problem shared by all models is the general difficulty in explaining a continuous distribution of relaxation times [17]. Alternative models [26,28,44], which don't invoke long-lived (> 1 s) electron traps, have, in turn, been suggested to account for this distribution of relaxation times.

**5. Fluctuating non-radiative rate models**: (**Figure 2f**) Frantsuzov and Marcus [44] have suggested that QD intermittency is a result of the fluctuations of the non-radiative recombination rate. Recombinations occur through the Auger-assisted excitation of deep surface states and followed by relaxation to the ground state. The trapping rate is then governed by the spectral diffusion of a second excited QD state ($1P_e$), which modulates the eventual non-radiative recovery of the system. This mechanism naturally explains a continuous distribution of relaxation times. However, it leaves unanswered several model constraints. In particular, it gives a -3/2 slope for both on and off distributions regardless of the threshold level. A modified version of the same model was suggested recently by Barnes and co-workers [26] to explain blinking suppression after ligand exchange. A similar model was developed by Barbara [28] to explain the apparent external electric field modulation of QD emission intensities.

### *Discussion and conclusion*

Single QDs, NWs and molecules all demonstrate universal emission intermittency over large timescales. Furthermore, a threshold analysis reveals truncated power-law distributions for both on-time and off-time probability densities. Explaining these two observations has been the cornerstone of much work in the field for the last 10 years. As shown above, available theoretical frameworks have successfully explained a large subset of experimental results. They include, but are not limited to *(a)-(g)*. However, no theory currently accounts for all the constraints. Although many experiments have been conducted, they have not been performed on the same quantum dot and in the same laboratory, making some theoretical tests less precise. More detailed intensity dependent as well as band edge excitation experiments are absent, but may be useful in better revealing a universal blinking mechanism.

### References

1. Cook, R. J. & Kimble H. J. Possibility of Direct Observation of Quantum Jumps. *Phys. Rev. Lett.* **54**, 1023-1026 (1985).

2. Moerner, W. E. & Orrit, M. Illuminating single molecules in condensed matter. *Science* **283**, 1670-1676 (1999).

3. Wustholz, K. L., Bott, E. D., Isborn, C. M., Li, X., Kahr, B. & Reid, P. J. Dispersive kinetics from single molecules oriented in single crystals of potassium acid phthalate, *J. Phys. Chem. C* **111**, 9146-9156 (2007).

4. Yeow, E. K. L., Melnikov, S. M., Bell, T. D. M., De Schryver, F. C. & Hofkens, J. Characterizing the fluorescence intermittency of photobleaching kinetics of dye molecules immobilized on a glass surface. *J. Phys. Chem. A*, **110**, 1726-1734 (2006).

5. Hoogenboom, J. P., van Dijk, E. M. H. P., Hernando, J., van Hulst, N. F. & Garcia-Parajo, M. F. Power-law–distributed dark states are the main pathway for photobleaching of single organic molecules. *Phys. Rev. Lett.* **95**, 097401 (2005).

6. Hoogenboom, J. P., Hernando, J., van Dijk, E. M. H. P., van Hulst, N. F. & Garcia-Parajo, M. F. Power-law blinking in the fluorescence of single organic molecules. *ChemPhysChem* **8**, 823-833 (2007).

7. Schuster, J., Cichos, F. & von Borczyskowski, C. Blinking of single molecules in various environments. *Opt. Spect.* **98**, 778-783 (2005).

8. Haase, M. et al. Exponential and power-law kinetics in single-molecule fluorescence intermittency. *J. Phys. Chem. B* **108**, 10445-10450 (2004).

9. Dickson, R. M., Cubitt, A. B., Tsien, R. Y. & Moerner, W. E. On/off blinking and switching behavior of single green fluorescent protein molecules. **Nature 388**, 355-358 (1997).

10. Vanden Bout, D. A., Yip, W. T., Hu, D., Fu, D. K. Swager, T. M. & Barbara, P. F. Discrete Intensity Jumps and Intramolecular Electronic Energy Transfer in the Spectroscopy of Single Conjugated Polymer Molecules. *Science* **277**, 1074-1077 (1997).

11. Nirmal, M., Dabbousi, O. B., Bawendi, M. G., Macklin, J. J., Trautman, J. K., Harris, T. D. & Brus, L.E. Fluorescence intermittency in single cadmium selenide nanocrystals. *Nature* **383**, 802-804 (1996).

12. Kuno, M., Fromm, D. P., Hamann, H. F., Gallagher, A. & Nesbitt, D. J. Nonexponential 'blinking' kinetics of single CdSe quantum dots: A universal power law behavior. *J. Chem. Phys.* **112**, 3117-3120 (2000).

13. Kuno, M., Fromm, D. P., Hamann, H. F., Gallagher, A. & Nesbitt, D. J. 'On/Off' fluorescence intermittency of single semiconductor quantum dots. *J. Chem. Phys.* **115**, 1028-1040 (2001).

14. Shimizu, K. T., et al. Blinking statistics in single semiconductor nanocrystal quantum dots. *Phys. Rev. B* **63**, 205316 (2001).

15. Pelton, M., Grier, D. G., Guyot-Sionnest, P. Characterizing quantum-dot blinking using noise power spectra. *Appl. Phys. Lett.* **85**, 819-821 (2004).

16. Pelton, M., Smith, G., Scherer, N. F. & Marcus, R. A. Evidence for a diffusion-controlled mechanism for fluorescence blinking of colloidal quantum dots. *Proc. Natl. Acad. Sci.* **104,** 14249-14254 (2007).

17. Zhang, K., Chang, H., Fu, A., Alivisatos, A. P. & Yang, H. Continuous distribution of emission states from single CdSe/ZnS quantum dots. *Nano Lett.* **6**, 843-847 (2006).

18. Schlegel, G., Bohnenberger, J., Potapova, I., Mews, A. Fluorescence decay time of single semiconductor nanocrystals. *Phys. Rev. Lett.* **88**, 137401 (2002).

19. Fisher, B. R., Eisler, H. J., Stott, N. E. & Bawendi, M.G. Emission intensity dependence and single exponential behavior in single colloidal quantum dot fluorescence lifetimes. *J. Phys. Chem. B* **108**, 143-148 (2004).

20. Chung, I. & Bawendi, M.G. Relationship between single quantum-dot intermittency and fluorescence intensity decays from collections of dots. *Phys. Rev. B* **70**, 165304 (2004).

21. Neuhauser, R. G., Shimizu, K. T., Woo, W. K., Empedocles, S. A. & Bawendi, M. G. Correlation between Fluorescence Intermittency and Spectral Diffusion in Single Semiconductor Quantum Dots. *Phys. Rev. Lett.* **85**, 3301-3304 (2000).

22. Fomenko V. & Nesbitt, D. J. Solution Control of Radiative and Nonradiative Lifetimes: A Novel Contribution to Quantum Dot Blinking Suppression. *Nano Lett.* **8**, 287-293 (2008).

23. Stefani, F. D., Zhong, X., Knoll, W., Han, M. & Kreiter, M. Memory in quantum-dot photoluminescence blinking. *New J. Phys.* **7**, 197 (2005).

24. Stefani, F. D., Zhong, X., Knoll, W., Kreiter, M. & Han, M. Quantification of photoinduced and spontaneous quantum-dot luminescence blinking. *Phys. Rev. B* **72**, 125304 (2005).

25. Hohng, S. & Ha, T. Near-complete suppression of quantum dot blinking in ambient conditions. *J. Am. Chem. Soc.* **126**, 1324-1325 (2004).

26. Hammer, N. I., Early, K. T., Sill, K., Odoi, M. Y., Emrick, T. & Barnes, M. D. Coverage-mediated suppression of blinking in solid state quantum dot conjugated organic composite nanostructures. *J. Phys. Chem. B* **110**, 14167-14171 (2006).

27. Fu, Y., Zhang, J. & Lakowicz, J. R. Suppressed blinking in single quantum dots (QDs) immobilized near silver island films (SIFs). *Chem. Phys. Lett.* **447**, 96-100 (2007).

28. Park, S. J., Link, S., Miller, W. L., Gesquiere, A. & Barbara, P. F. Effect of electric field on the photoluminescence intensity of single CdSe nanocrystals. *Chem. Phys.* **341**, 169-174 (2007).

29. Wang, S., Querner, C., Emmons, T., Drndić, M. & Crouch, C. H. Fluorescence blinking statistics from CdSe core and core-shell nanorods. *J. Phys. Chem. B* **110**, 23221-23227 (2006).

30. Protasenko, V. V., Gordeyev, S. & Kuno M. Spatial and intensity modulation of nanowire emission induced by mobile charges. *J. Am. Chem. Soc.* **129**, 13160-13171 (2007).

31. Protasenko, V. V., Hull, K. L. & Kuno M. Disorder-induced optical heterogeneity in single CdSe nanowires. *Adv. Mater.* **17**, 2942-2949 (2005).

32. Glennon, J. J. Tang, R., Buhro, W. E. & Loomis, R. A. Synchronous photoluminescence intermittency (blinking) along whole semiconductor quantum wires. *Nano Lett.* **7**, 3290-3295 (2007).

33. Kadanoff, L. P. Static Phenomena near critical points – Theory and experiment. *Rev. Mod. Phys.* **39,** 395-431 (1967).

34. Cichos, F., von Borczyskowski, C. & Orrit, M. Power-law intermittency of single emitters. *Curr. Op. Coll. Int. Scie.* **12**, 272-284 (2007).

35. Gomez, D. E., Califano, M. & Mulvaney, P. Optical properties of single semiconductor nanocrystals. *Phys. Chem. Chem. Phys.* **8**, 4989-5011 (2006).

36. Margolin, G., Protasenko, V., Kuno, M. & Barkai, E. Power law blinking quantum dots: Stochastic and physical models. *Adv. Chem. Phys.* **133,** 327-356 (2006).

37. Tang, J. & Marcus, R. A. Single particle vs. Ensemble Average: From Power-law Intermittency of a Single Quantum Dot to Stretched Exponential Fluorescence Decay of an Ensemble. *J. Chem. Phys.* **123**, 054704 (2005).

38. Sher, P. H. et al. Power law carrier dynamics in semiconductor nanocrystals at nanosecond timescales. *Appl. Phys. Lett.* **92**, 101111 (2008).

39. Margolin, G. & Barkai, E. Nonergodicity of blinking nanocrystals and other Levy-walk processes. *Phys. Rev. Lett.* **94**, 080601 (2005).

40. Bianco, S., Grigolini, P. & Paradisi, P. *Fluorescence intermittency in blinking quantum dots: Renewal or slow modulation? J. Chem. Phys.* **123**, 174704 (2005).

41. Efros, A. L. & Rosen, M. Random telegraph signal in the photoluminescence intensity of a single quantum dot. *Phys. Rev. Lett.* **78**, 1110-1113 (1997).

42. Verberk, R., van Oijen, A. M. & Orrit, M. Simple model for the power-law blinking of single semiconductor nanocrystals. *Phys. Rev. B* **66**, 233202 (2002).

43. Tang, J. & Marcus, R. A. Diffusion-controlled electron transfer processes and power-law statistics of fluorescence intermittency of nanoparticles. *Phys. Rev. Lett.* **95**, 107401 (2005).

44. Frantsuzov, P. A. & Marcus, R. A. Explanation of quantum dot blinking without the long-lived trap hypothesis. *Phys. Rev. B*. **72**, 155321 (2005).

45. Krauss, T. D. & Brus, L. E. Charge, Polarizability, and Photoionization of Single Semiconductor Nanocrystals. *Phys. Rev. Lett.* **83**, 4840-4843 (1999).

46. Janko, B. & Ambegaokar V. Parity fluctuations between Coulomb Blockaded superconducting islands. *Phys. Rev. Lett.* **75**, 1154-1157 (1995).

Correspondence should be addressed to B. J. (bjanko@nd.edu).

**Acknowledgements**: The authors would like to acknowledge the support of NSF, ONR and DOE-BES.

**Competing Interests:** The authors declare that they have no competing interests.

**DOI: 10.1038/nphys1001**

URL: http://www.nature.com/nphys/journal/v4/n7/full/nphys1001.html

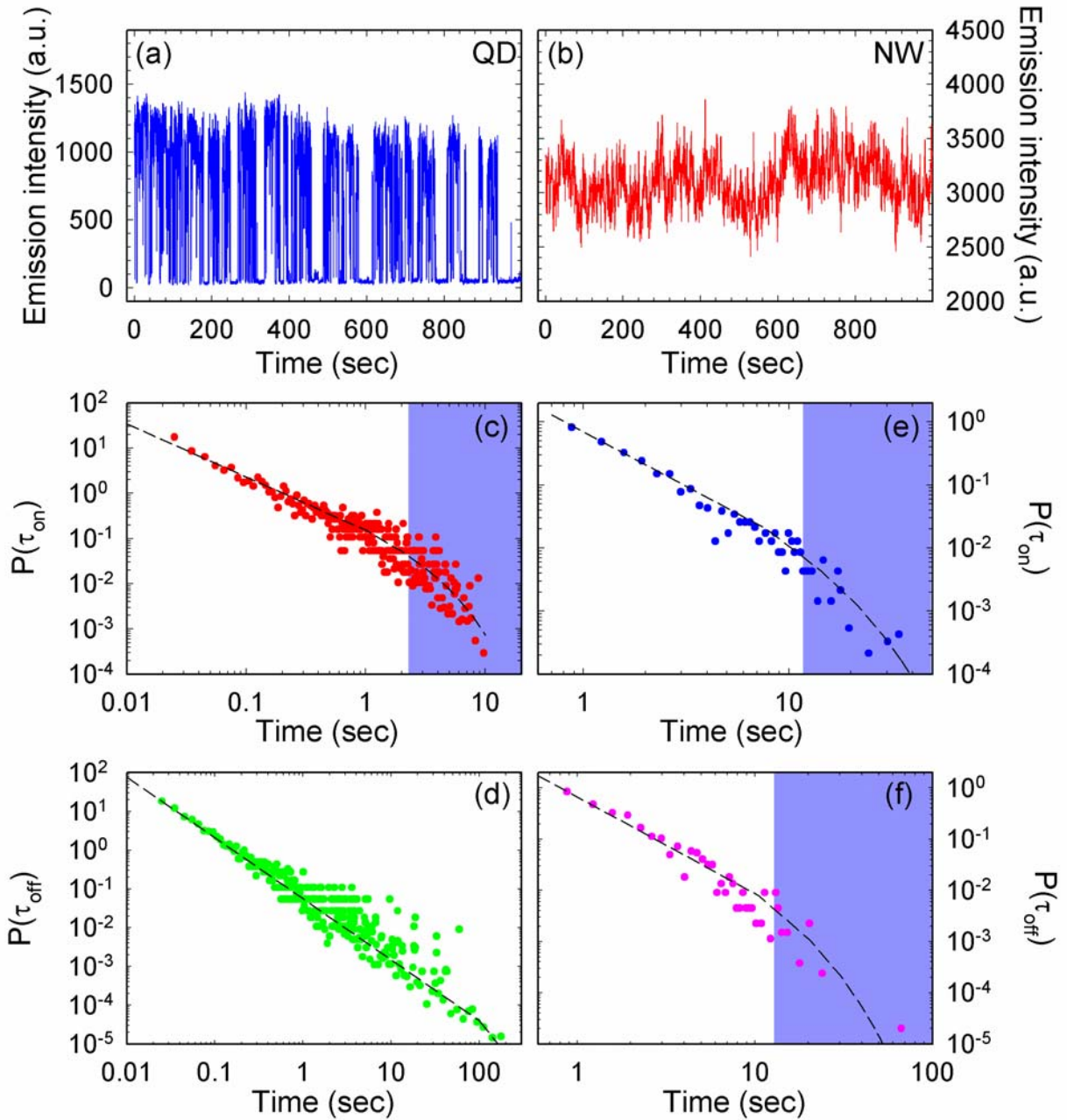

**Figure 1**. **Time distributions for intermittency.** Emission trajectory from a single (a) quantum dot and (b) nanowire. Corresponding QD log-log plot of the (c) on-time and (d) off-time probability density. Analogous log-log plot of the NW (e) on-time and (f) off-time probability density.

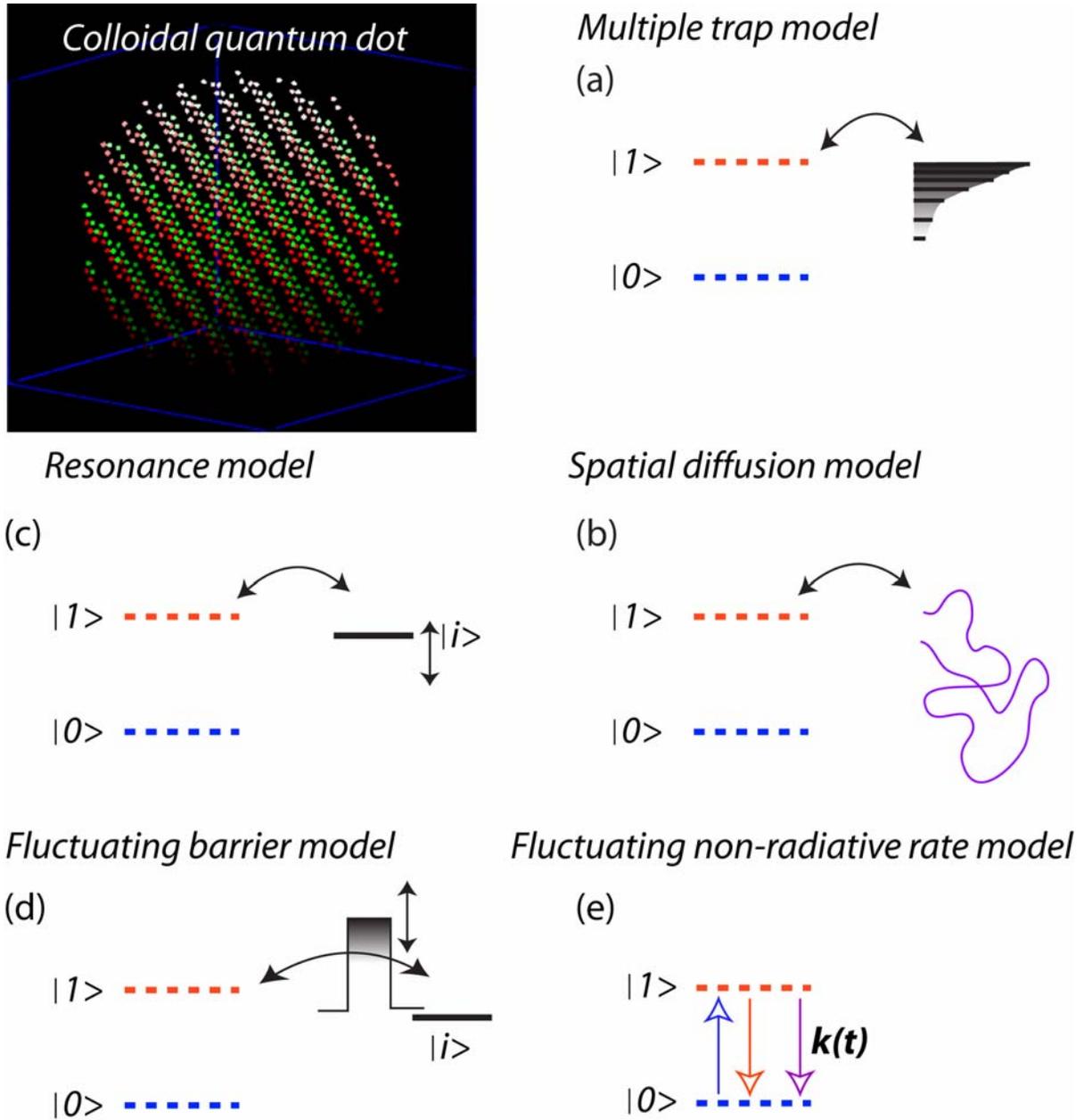

**Figure 2**. **Schematic diagrams of blinking models**. (a) 3D structure of a CdSe colloidal quantum dot. Red (green) dots represent Cd (Se) atoms, respectively; (b) An electron jumps from the QD excited state to one of the multiple traps and returns. (c) The electron jumps to/from the trap when it is in resonance with the exited state. (d) The ejected electron performs 3D diffusion in the surrounding and returns. (e) The tunneling barrier between the QD and the trap state randomly changes due to electron jump. (f) The non-radiative relaxation rate of the excited state (purple arrow) carries out fluctuations correlated over long time intervals.